\documentclass[conference]{IEEEtran}

\ifCLASSINFOpdf
  \usepackage[pdftex]{graphicx}
  \DeclareGraphicsExtensions{.pdf,.jpeg,.png}
\else
\fi

\begin{document}
%
\title{Thermoelectric properties of disordered graphene antidot devices}


\author{\IEEEauthorblockN{Tue Gunst\IEEEauthorrefmark{1}\IEEEauthorrefmark{4}, Jing-Tao L\"u\IEEEauthorrefmark{1}\IEEEauthorrefmark{4}, Troels Markussen\IEEEauthorrefmark{2}\IEEEauthorrefmark{4}, Antti-Pekka Jauho\IEEEauthorrefmark{1}\IEEEauthorrefmark{3}\IEEEauthorrefmark{4} and Mads Brandbyge\IEEEauthorrefmark{1}\IEEEauthorrefmark{3}\IEEEauthorrefmark{4}}

\IEEEauthorblockA{\IEEEauthorrefmark{3}Center for Nanostructured Graphene (CNG)}
\IEEEauthorblockA{\IEEEauthorrefmark{4}Technical University of Denmark DK-2800 Kgs. Lyngby, Denmark}
\IEEEauthorblockA{\IEEEauthorrefmark{1}Department of Micro- and Nanotechnology (DTU Nanotech), Email: Tue.Gunst@nanotech.dtu.dk}
\IEEEauthorblockA{\IEEEauthorrefmark{2}Center for Atomic-scale Materials Design (CAMD), Department of Physics}}



\maketitle

\begin{abstract}
We calculate the electronic and thermal transport properties of devices based on finite graphene antidot lattices (GALs) connected to perfect graphene leads. We use an atomistic approach based on the $\pi$-tight-binding model, the Brenner potential, and employing recursive Green's functions. We consider the effect of random disorder on the electronic and thermal transport properties, and examine the potential gain of thermoelectric merit by tailoring of the disorder. We propose several routes to optimize the transport properties of the GAL systems. Finally, we illustrate how quantum thermal transport can be addressed by molecular dynamics simulations, and compare to the Green's function results for the GAL systems in the ballistic limit.
\end{abstract}
%
\IEEEpeerreviewmaketitle

\section{Introduction}
Thermal management in nanoelectronics is a topic of great importance\cite{balandin_thermal_2011}, and it is relevant to consider materials for
thermo-electric cooling at the nanoscale. Although graphene is an excellent heat conductor, its thermoelectric properties seem to be highly tunable in nanostructured graphene, such as graphene antidot lattices (GALs)\cite{gunst_thermoelectric_2011}. GALs have been proposed as a flexible platform for creating a semiconducting
material\cite{pedersen_graphene_2008,fuerst_electronic_2009,Petersen_ACSNano_2011,Ouyang_ACSNano_2011} and the regular nanoperforation of a graphene layer may be a solution for making graphene a versatile material for electronics as well as thermal management.

Several groups have developed fabrication techniques for creating this nanostructured system, e.g. by electron beam lithography\cite{eroms_weak_2009,Beg_NanoLett_2011} or block copolymer lithography\cite{kim_fabrication_2010,bai_graphene_2010}.
The latter method results in a highly regular hexagonal lattice of holes. In another experiment hexagonal holes were obtained with very pure edge chirality by anisotropic etching\cite{krauss_raman_2010}.
Various types of disorder play an important role for transport properties.
Edge disorder in nanoribbons has previously been theoretically predicted to increase the thermoelectric efficiency\cite{sevincli_enhanced_2010} and widen the band gap distribution of GALs with periodic disorder\cite{jippo_theoretical_2011}.
Recent experiments have managed to control the concentration of the $^{13}\rm{C}$ isotope in both amount and space\cite{li_evolution_2009} and it has also been possible to measure the thermal conductivity by Raman spectroscopy\cite{chen_thermal_2012}. Several theoretical predictions have been made on the isotope effect in graphene\cite{mingo_cluster_2010,zhang_isotope_2010}, nanoribbons\cite{hu_tuning_2010,jiang_isotopic_2010,huang_simulation_2010} and carbon nanotubes\cite{yamamoto_universality_2011}.

Here we use atomistic modelling to address how structural and isotope disorder can reduce the thermal conductance of GAL devices.

\section{The role of disorder in GALs}
The electronic and phonon transport properties are calculated from the energy-dependent transmission functions, $\mathcal{T}_e$ and $\mathcal{T}_{ph}$, using a Landauer-type formula.
The Landauer formula for the electrons reads
\begin{eqnarray}
I_e = \frac{2e}{\hbar}\int \frac{dE}{2\pi}\mathcal{T}_e(E) [n_F(E,\mu_L)-n_F(E,\mu_R)]\,,\label{eqn:Landauer}
\end{eqnarray}
where $n_F(E,\mu_{L/R})$ is the Fermi-Dirac distribution at the chemical potential of the left/right lead.
Within the linear-response limit we can consider variations with changes in the chemical
potential, $\mu$, e.g. by doping or gating of the graphene system, by evaluated the following moments of the electronic transmission,
\begin{equation}
L_n(\mu) = \frac{2}{\hbar}\int\frac{dE}{2\pi}(E-\mu)^n\mathcal{T}_e(E)\left(-\frac{\partial n_F}{\partial E}\right).\label{eqn:Ln}
\end{equation}
They relate the electronic current and the electron heat current, in the linear response regime. Thus,
we have\cite{sivan_multichannel_1986}
the electrical conductance $G_e(\mu)=\frac{\partial I}{\partial V}=e^2L_0$, the electron thermal conductance
$\mathcal{\kappa}_e(\mu) = \left[L_2-\frac{L_1^2}{L_0}\right]/T$, and the Seebeck coefficient $S(\mu) = \frac{\Delta V}{\Delta T}|_{I_e=0} = \frac{L_1}{eL_0T}$.
For phonons the Landauer formula takes an analogous form and within linear response the thermal conductance is given by
\begin{eqnarray}
\kappa_{ph} = \int_{0}^{\infty}d\omega \frac{\left(\hbar \omega\right)^2}{2\pi k_B T^2}
\mathcal{T}_{ph}(\omega) \frac{e^{\frac{\hbar \omega}{k_B T}}}{(e^{\frac{\hbar \omega}{k_B T}}-1)^2}\,.
\end{eqnarray}
From these we can evaluate the thermoelectric figure of merit,
\begin{eqnarray}
ZT = \frac{S^2 G_e T}{\kappa_{ph}+\kappa_e}
\label{eqn:ZT}\,.
\end{eqnarray}
\begin{figure}[h!tbp]
\centering
{\includegraphics[width=0.4\paperwidth]{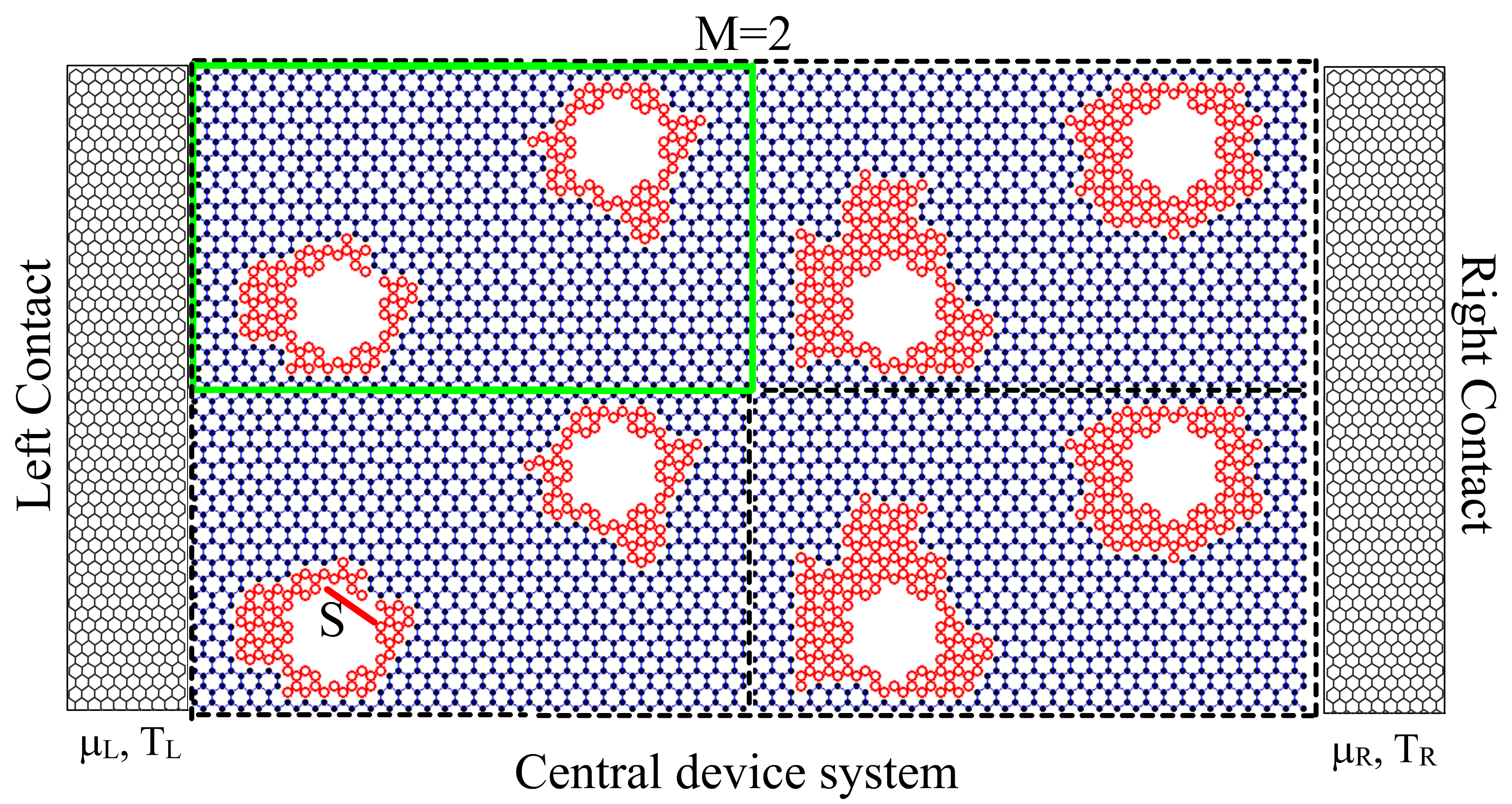}}
\caption{System setup and the computational rectangular unit cell (green rectangle). Two graphene leads are connected by the finite GAL of length 2 ($M=2$) corresponding to 4 holes in the direction of transport. The depicted system is based on a $\{L=10,S=3\rm{arm}\}$ GAL, where we use the nomenclature from \cite{gunst_thermoelectric_2011}. The red atoms are randomly removed during the generation of edge noise. The unit cell is $10$ four-atom armchair units wide. We apply periodic boundary conditions in the transverse direction.}
\label{fig:System}
\end{figure}
We consider only disorder in the transport direction in GAL structures with periodicity in the direction perpendicular to transport, see Fig.~\ref{fig:System},
and employ averaging of the transmission functions over a dense $k_\perp$-point sampling.
Using the Brenner potential\cite{brenner_empirical_1990} we perform structural relaxation, obtain the dynamical matrix,
and then calculate the electronic Hamiltonian using the $\pi$-tight-binding model for the device region of a given length between pristine graphene leads.
Based on this we calculate the $k$-averaged $\mathcal{T}_e$ and $\mathcal{T}_{ph}$ by an atomistic Green's function method, which is efficient for long systems with different kinds of disorder\cite{markussen_electron_2009,esfarjani_thermoelectric_2006}.
Note that we assume the device to be free-standing, and neglect all effects of electron-phonon interaction such as phonon-drag.
Later, we comment on how anharmonic effects may be included using molecular dynamics.
In the remaining of this section we apply this method to three types of disorder; edge disorder, position disorder, and the isotope effects.

\subsection{Lattice disorder}
\begin{figure}[h!tbp]
\centering
{\includegraphics[width=0.37\paperwidth]{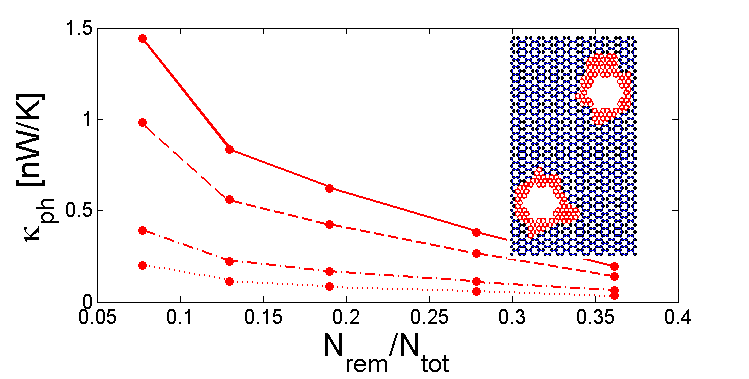}}
{\includegraphics[width=0.37\paperwidth]{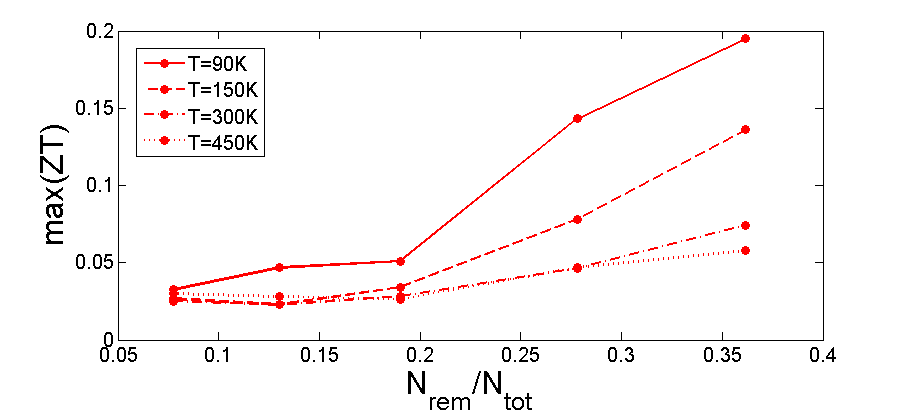}}
\caption{
Edge disorder in a GAL system with 5 unit cells along the device, corresponding to a length of $380$\AA. The system is based on a \{10,2.3arm\} GAL (with a width of $43$\AA) and we increase the number of atoms removed at random positions at the edge (the red atoms in the insert, showing one unit cell at $N_{rem}/N_{tot}=0.19$). The ratio of removed atoms to unit cell size is changed through the interval previously considered for perfect antidots\cite{gunst_thermoelectric_2011}. The resulting edges are highly disordered and can be taken as a worst case scenario (e.g. by vapor deposition of carbon atoms on a surface).}
\label{fig:EdgeDisorder}
\end{figure}

In Fig.~\ref{fig:EdgeDisorder}, we show the outcome of transport simulations for various GALs with random edge disorder.
The system is constructed from a sequence of GAL unit cells (with two holes) in which a chosen number of atoms are
randomly removed at the edge one by one. We start from a very small \{10,2.3arm\}-GAL, and increase the hole size in this
random fashion until it is comparable to the \{10,6arm\}, which have a $max(ZT)\approx 0.25$ at 300K\cite{gunst_thermoelectric_2011}.
Here, $max(ZT)$ refers to the maximal occurring $ZT$ within a gate bias range of $[-1.5;1.5]$eV.
We do not allow for C-atoms with a single nearest neighbor, since these are energetically less favorable.
Regarding edge disorder the decrease of thermal conductance outweights
the decrease in the power factor ($S^2 G_e$), as can be seen in the increase in $ZT$
with the number of removed atoms in Fig.~\ref{fig:EdgeDisorder}. However, the effect for disordered antidots is much smaller as compared to the effect for hole-size for edges with pure chirality\cite{gunst_thermoelectric_2011}. Therefore, one should preferable avoid the edge disorder during fabrication if one seeks to optimize thermoelectric applications. We note that the kinetics of the system or electronic current\cite{jia_controlled_2009} may stabilize pure edges to some extent.

\begin{figure}[h!tbp]
\centering
{\includegraphics[width=0.37\paperwidth]{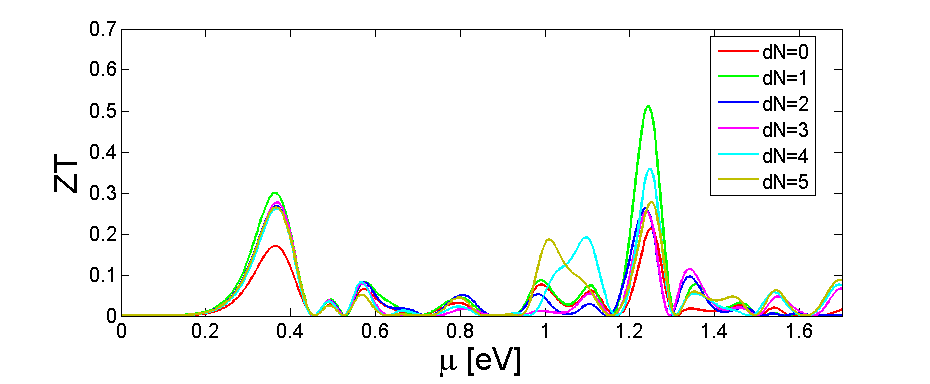}}
\caption{Position disorder is here studied by increasing an interface mismatch between two GAL regions. Increasing the position disorder tends to reduce both the thermal conductance and also the power factor except at a very low degree of disorder. The system has a length of 380\AA ($M=5$) and is based on a $\{10,6arm\}$ GAL.}
\label{fig:PositionDisorder}
\end{figure}

Next, we examine disorder in the hole lattice positions by randomly shifting a sequence of 2-hole unit-cells along the transverse
direction by $dN$ (armchair unit periodicity). The resulting figure of merit is given in Fig.~\ref{fig:PositionDisorder}.
The full cell is $L=10$ armchair units wide, and we examine $dN=1,2,..5$ corresponding to increasing lateral disorder.
The electronic conductance is almost unchanged at a low degree of position disorder ($dN=1$), and the power factor can even initially increase at a low degree of position disorder, before it eventually decreases. Meanwhile, the thermal conductance decreases with increasing $dN$. Therefore, especially a weak degree of position disorder increases $ZT$ to above $0.5$. For perfect GALs with a high antidot density the electronic $ZT$ (setting $\kappa_{ph}=0$) is not much larger at the chemical potential of the largest peak. For low $\mu$ the $\kappa_{ph}$ dominates, and therefore it is the first $ZT$ peak which is increased by the disorder.
On the other hand, the highest peak at higher $\mu$ mainly follows the change in power factor.

\subsection{Isotope disorder}
\begin{figure}[h!tbp]
\centering
{\includegraphics[width=0.37\paperwidth]{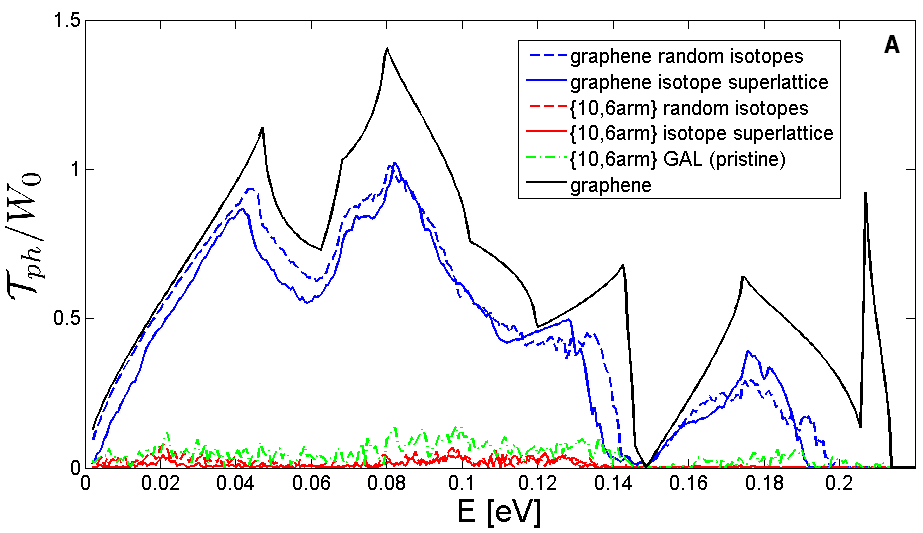}}
{\includegraphics[width=0.37\paperwidth]{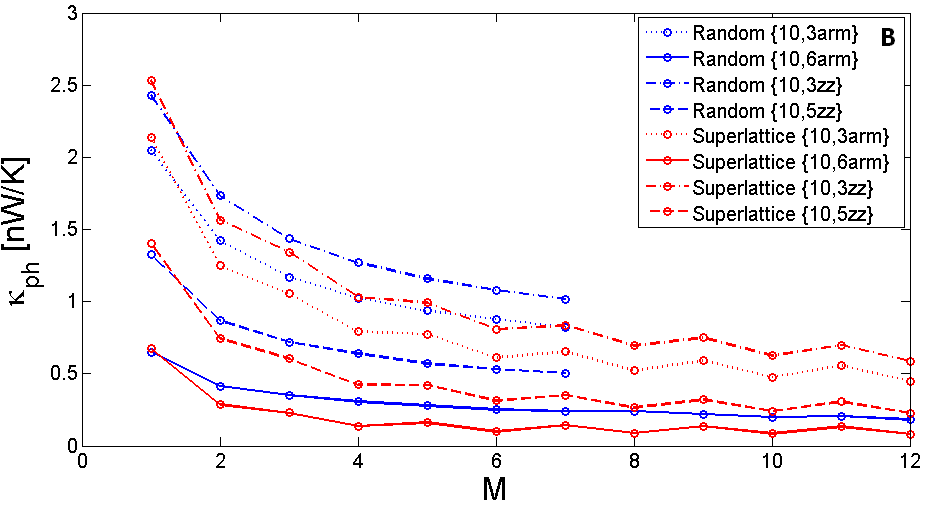}}
{\includegraphics[width=0.37\paperwidth]{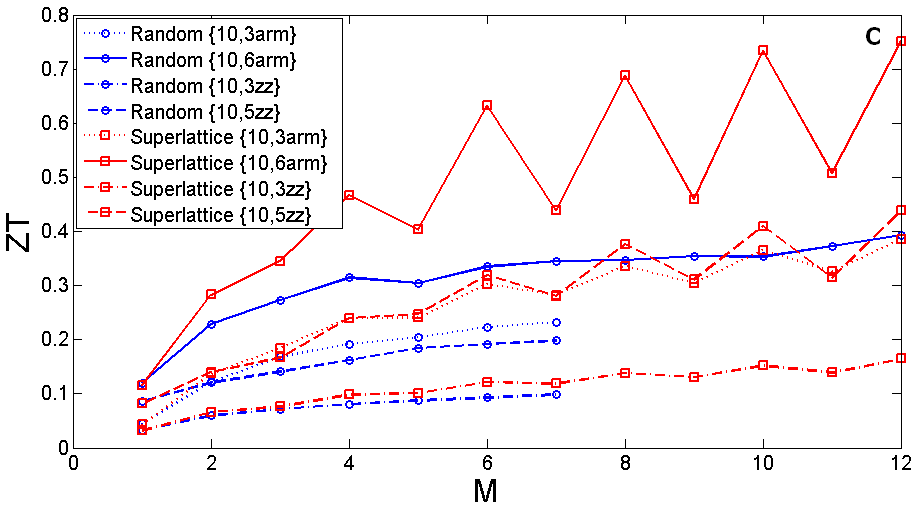}}	
\caption{The effect of isotopes is considered for a long system ($M=12$). The random isotopes of $^{13}\rm{C}$ atoms are distribution at random positions at each $M$ with a concentration of $50\%$. For the supercell pattern every second unit cell consist of $^{13}\rm{C}$ isotopes. The temperature is 300K.}
\label{fig:Isotopes}
\end{figure}

Now we address the effect of isotopes positioned at random sites or in a pattern changing with every 2nd 2-hole supercell (Fig.~\ref{fig:Isotopes}).
The effect of isotopes on the thermal conductivity of graphene has recently attracted much attention due to the advances in graphene synthesized by chemical vapour deposition (CVD). By using sequential input of $^{13}\rm{CH}_4$ and $^{12}\rm{CH}_4$ several groups have managed to fabricate graphene with regions of different isotopes\cite{li_evolution_2009}. The local thermal conductivity of the regions with different isotopes has been measured by Raman spectroscopy\cite{chen_thermal_2012}. The isotope effect has been addressed theoretically for graphene\cite{mingo_cluster_2010,zhang_isotope_2010} and nanoribbons\cite{hu_tuning_2010}. Inspired by this we here consider how the isotopes reduce the thermal conductance of GAL devices.
In Fig.~\ref{fig:Isotopes}A we show $\mathcal{T}_{ph}$ (scaled by cell-width) for the longest system considered ($M=12$ two-hole unit cells). We clearly see that isotope scattering give rise to a significant reduction in the thermal transmission. Comparing the reduction from the random isotope distribution to that of a pattern (alternating isotopes for every 2-hole supercell), it is clear that the "clustering" of isotopes leads to an extra reduction for especially the acoustic modes in the GAL region. Furthermore, we see that the difference will be less important at higher temperatures ($>300K$ used here). In Fig.~\ref{fig:Isotopes}b we consider the thermal conductance for two GAL systems with different hole shape and size, as a function of length of GAL-device. For comparison, the thermal conductance is found to be 6.1nW/K for a pristine graphene device of the same width, and 0.41nW/K, 1.17nW/K, 0.76nW/K and 1.38nW/K for ideal \{10,6arm\}, \{10,3arm\}, \{10,9zz\} and \{10,3zz\} GALs, with length $M>5$, respectively.
With random isotopes the thermal conductance of the $M=12$-\{10,6arm\}-device is reduced by $55\%$, compared to $88\%$ with the ordered isotope pattern. For devices of longer lengths we enter the diffusive region and cannot neglect phonon-phonon scattering as considered here.

Since isotopes only affect the thermal properties we get a significant increase in $ZT$, especially at low gate bias where thermal transport is dominated by phonons. A peak $max(ZT)\approx$ 0.8 is observed for the supercell pattern of isotopes at 300K, compared to $max(ZT)\approx$ 0.26, 0.17, 0.07 and 0.13 for the ideal \{10,6arm\}, \{10,3arm\}, \{10,9zz\} and \{10,3zz\} GALs without isotopes. The oscillations in $ZT$ with $M$ stems from the choice of patterning where we only have $50-50\%$ isotopes for even $M$. With random isotopes we can exceed $max(ZT)\approx$ 0.4 for the \{10,6arm\} lattice. This illustrates that one can benefit from a further reduction of the heat conductance in GAL devices to increase the thermoelectric efficiency.

\section{A molecular dynamics approach to quantum thermal transport}
Finally, we will briefly discuss how the elastic quantum phonon transport can be obtained from generalized Langevin molecular dynamics(MD), as an alternative to the Green's function method. The method was proposed by Wang \textit{et al.}\cite{wang_quantum_2007}, and is a promising method for handling both low temperature quantum freeze-out of phonons, as well as the anharmonic interaction beyond perturbation theory.
Both of these effects need to be included for graphene at 300K for devices longer than $\approx 600$\AA\cite{wang_molecular_2009}.
Anharmonic effects are clearly a very challenging problem for larger nanostructures such as GALs, where the role of the nanostructuring on the thermal properties needs clarification. 

We solve the generalized Langevin equation for the mass-normalized displacements,
\begin{eqnarray}
\vec{\ddot{u}}_s(t) &=& \vec{F}_{con}-\int_{t_0}^{t}\mathbf{\Phi}(t-t')\cdot \vec{\dot{u}}_{s}(t')dt' + \tilde{\varepsilon}_{ph}(t)\,,\label{PhononLangevin}
\end{eqnarray}
by using the velocity-Verlet algorithm. The forces on the right hand side are from the full Brenner potential, the frictional damping force and the quantum noise. 
The two latter are due to the connection of the device to the two thermal baths (each lead). 
The essential variable in quantum thermal transport is the coupling matrix obtained from the retarded phonon self energy, $\mathbf{\Gamma}_b(\omega)=i(\mathbf{\Pi}^r(\omega)-\mathbf{\Pi}^a(\omega))$. This object describes how much heat is absorbed(emitted) to(from) the surroundings. The non-Markovian damping kernel is obtained from $\mathbf{\Gamma}_b$ as\cite{wang_quantum_2007}:
\begin{eqnarray}
\mathbf{\Phi}(\tau)=\frac{\Delta \omega}{\pi}\sum_{k=1}^{N_k}\left[\frac{\Gamma_b(\omega_k)}{\omega_k-i \epsilon_{ad}} e^{-i \omega_k \tau-\epsilon_{ad} \tau}+c.c.\right]\,.
\end{eqnarray}
Where $\tau=t-t'$ is the relative time so that the frictional damping force includes the memory at time $t$ of a displacement at time $t'$.
We choose a discrete time step $\Delta t$ for a $N_k$ step long MD simulation (with corresponding $\Delta \omega$ and discrete $\omega_k$ in frequency space).
To improve convergence and computation-time we include an artificial damping, $\epsilon_{ad}$, to reduce the memory kernel length, $n_{cut}$. 
We can from this calculate an effective imaginary part of the phonon self energy including this extra damping,
\begin{eqnarray}
\mathbf{\Gamma}_b^{eff}(\omega)=\sum_{i=1}^{n_{cut}}\omega \Delta t \mathbf{\Phi}(\tau_i) e^{i \omega \tau_i}
\end{eqnarray}
From this we can calculate the noise spectral power density from the quantum fluctuation-dissipation theorem:
\begin{eqnarray}
\mathbf{S}_f^{ph}(\omega) &=& \hbar\left(n_B(\omega) +\frac{1}{2}\right) \mathbf{\Gamma}_b^{eff}(\omega)
\end{eqnarray}
We have implemented this method of generating a colored noise sequence and show several tests of the outcome with a large artificial damping below. The aim is to find out how accurate one has to represent the self-energy in terms of memory length and artificial damping to get results consistent with NEGF calculations.
\begin{figure}[h!tbp]
\centering
{\includegraphics[width=0.38\paperwidth]{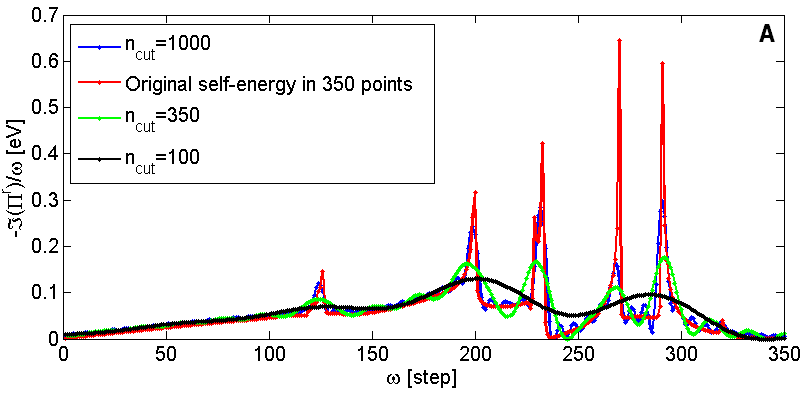}}
{\includegraphics[width=0.38\paperwidth]{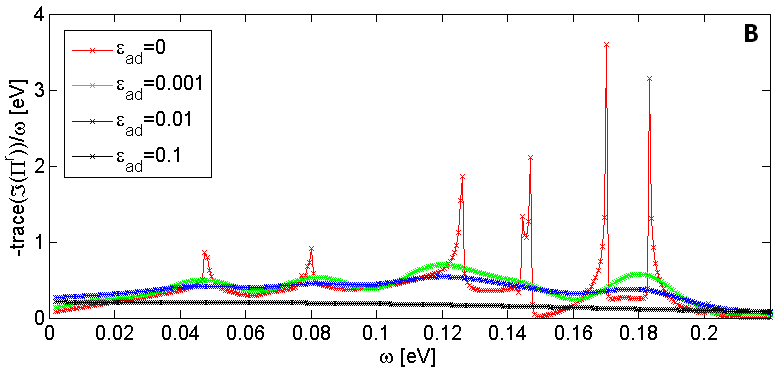}}
{\includegraphics[width=0.43\paperwidth]{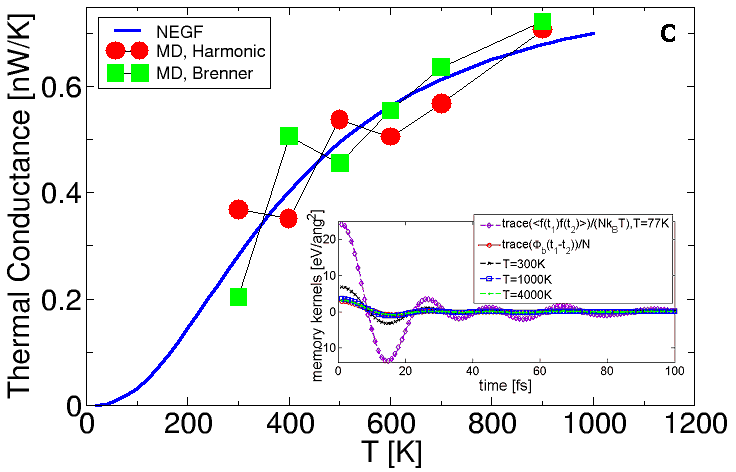}}
\caption{A: Effect of memory length in the effective self energy used in MD. B: Effect of artificial damping in the effective self energy. C insert: Temperature dependence of the noise correlation function compared to the damping kernel. C: Thermal conductance from nonequalibrium MD with a temperature difference of $10\%$ compared to the NEGF result for a \{10,3arm\} GAL with $M=1$. $\epsilon_{ad}=0.05$eV, $\delta t = 0.5$fs and $n_{cut}=350$ is used.}
\label{fig:MD}
\end{figure}
In Fig.~\ref{fig:MD} we show the imaginary part of the frequency dependent phonon bath self energy, and the noise correlation function. The first two figures (A and B) show how, even if we shorten the memory length (which in principle is as long as the simulation itself) to 100 time steps and increase the artificial damping to 0.1eV, the selfenergy is still approximated quite well. In the insert in Fig.~\ref{fig:MD}C we have calculated the noise correlation function (noise product at two different times averaged over $10^4$ starting points), which classically should equal $k_B T \mathbf{\Phi}(t-t')$. This is not the case for graphene at 300K. As can be seen from the figure one has to go above 1000K for the classical fluctuation-dissipation theorem to be a good approximation. The reason for this is the quantum fluctuations, which give rise to a larger noise and kinetic energy below the Debye temperature, and the Bose-distribution of the phonons. Furthermore, we see from the last figure that MD with quantum heat baths gives results for the thermal conductance consistent with NEGF for short devices even at low temperatures.

\section{Conclusion}
We have considered electronic and thermal transport properties of finite GAL structures with disorder.
We conclude that disorder can be a limiting factor for thermoelectric applications of GALs. A high degree of disorder either due to lattice irregularity or the lack of pure edge chirality will degrade the overall thermoelectric performance. However, the trend that an increased antidot density enhances the thermoelectric efficiency still survives. The presence of isotopes is found to drastically reduce the thermal conductance for GALs. Both in the case of lattice disorder and isotope mass-variation we have shown that disorder can be used to greatly enhance thermoelectric performance. In particular, if lattice disorder is kept at a minimum, and isotopes are changed in patterns on a length scale that affect most phonon modes, one can benefit from a very large reduction of the thermal conductance in such devices to obtain a good thermoelectric efficiency. Finally, we have illustrated how the harmonic quantum thermal transport can be obtained from generalized Langevin-MD simulations with quantum heat baths, yielding results similar to those obtained with the Green's-function/Landauer approach. The MD approach is, in principle, able to address the effects of anharmonicity in thermal transport of nanostructured graphene.


\section*{Acknowledgment}
The Center for Nanostructured Graphene is sponsored by
the Danish National Research Foundation. We thank the DCSC for computational resources.



%

\bibliographystyle{unsrt}


\end{document}